# Giant self-driven exciton-Floquet signatures in time-resolved photoemission spectroscopy of MoS₂ from time-dependent GW approach

Y.-H. Chan[1,2,3*], Diana Y. Qiu[1,2,4*], Felipe H. da Jornada[1,2,5,6*], and Steven G. Louie[1,2*]

**[1]Department of Physics, University of California, Berkeley, CA 94720-7300, USA**

**[2]Materials Sciences Division, Lawrence Berkeley National Laboratory, Berkeley, CA 94720, USA**

**[3]Institute of Atomic and Molecular Sciences, Academia Sinica, Taipei 10617, Taiwan and Physics Division, National Center of Theoretical Sciences, Taiwan**

**[4]Department of Mechanical Engineering and Materials Science, Yale University, New Haven, CT 06520**

**[5]Department of Materials Science and Engineering, Stanford University, Stanford, CA 94305, USA**

**[6]Stanford PULSE Institute, SLAC National Accelerator Laboratory, Menlo Park, California 94025, USA**

*Corresponding author – email:yanghao@gate.sinica.edu.tw, diana.qiu@yale.edu, jornada@stanford.edu, sglouie@berkeley.edu

## Abstract

Time-resolved, angle-resolved photoemission spectroscopy (TR-ARPES) is a one-particle spectroscopic technique that can probe excitons (two-particle excitations) in momentum space. We present an *ab initio*, time-domain GW approach to TR-ARPES and apply it to monolayer MoS₂. We show that photoexcited excitons may be measured and quantified as satellite bands, as well as leading to the renormalization of the quasiparticle bands. These features are explained in terms of an exciton-Floquet phenomenon induced by an exciton time-dependent bosonic field, which is orders of magnitude stronger than laser field induced Floquet bands in low-dimensional semiconductors. Our findings open a door to understanding the behavior of optical-field driven materials.

***Introduction.*** Correlated electron-hole pair excitations, or excitons, typically dominate optical responses in semiconductors, both in terms of optical strength and new features in the spectrum going beyond the independent particle picture especially in reduced-dimensional systems. These correlated two-particle excitations are important in many opto-electronic applications and lead to exotic phenomena such as exciton condensates and singlet fission of excitons. Excitons are most often studied by optical experiments, in which low-energy excitons can be identified as discrete energy states in the band gap. Further characterization of excitons may be obtained on how these states respond to external parameters such as temperature, carrier doping, strain, and additional external fields.[1] In particular, pump-probe type optical experiments are able to provide essential information about how different exciton levels couple.[2, 3] In such experiments, a pump light is used to dress selected energy levels and a probe light can then be used to detect the energy level shifts of the dressed exciton states. A careful theoretical analysis in general is necessary to provide the characterization.

Time-resolved, angle-resolved photoemission spectroscopy (TR-ARPES) is a well-established tool for studying the dynamics of electron response. The resolution of events on a femtosecond or attosecond scale allows for the investigation of nonequilibrium phenomena [4], such as melting of phases [5, 6], ultrafast band gap renormalization [7, 8], and transient carrier dynamics [9–13]. In recent years, the idea of using TR-ARPES to study excitons was proposed [14-17] and, more recently, realized experimentally [18-22]. These TR-ARPES experiments have been able to attain the necessary low temperature, spot size, and momentum resolution to measure some expected spectroscopic signatures of excitons qualitatively: satellite features with dispersion relation follows a replica dispersion of the valence band, which present even at time significantly after the passing of the external driving pump field [18,19].

Several studies have addressed the possibility of visualizing exciton features in TR-ARPES theoretically using model Hamiltonians [14-17, 23]. Quasi-equilibrium theory has also been applied to model a mixture of free carrier and bound excitons, where the spectral function shows a satellite structure due to excitons [24-26]. In these quasi-equilibrium models [24-26], excitons or carriers are assumed to be in equilibrium (taking advantage of the time-scale difference between electron-electron scattering and electron scattering due to other degree of freedoms), and hence cannot address the fast electron dynamics in the femtosecond time scale. Recently, time-dependent simulations of model systems suggest interesting features in nonequilibrium samples, such as signatures of excitonic insulator-like features [26,27]. However, a time-dependent first-principles simulation of TR-ARPES for real materials that includes electron-hole interactions remains a major challenge and is currently under active development [28,29].

Here, we develop a first-principles GW approach in the time domain [30,31] to compute the TR-ARPES in a pump-probe setup and apply it to the specific case of a monolayer $MoS_2$. We show that the detailed properties of the optically bright excitons may be obtained, resulting from a modification of the spectral function (*i.e.*, changes in the self-energy) of the measured hole, owing to the formation of excitons by the pump light. Features in the TR-ARPES occur as satellite bands at energies of the valence band plus an exciton excitation energy, which yields the **k**-space wavefunction of the exciton. Moreover, under appropriate pump conditions in resonance with exciton excitations, the dispersion at the valence band maximum is renormalized into a camelback-like shape, which has previously been considered as a signature for a driven exciton insulator. We explain both the presence of the satellite band and the renormalization of the highest valence band shape in terms of a self-driven exciton-Floquet phenomenon induced by the photoexcited excitons' time-dependent bosonic field, which leads to a time-dependent single-particle self-energy coupling valence to conduction states — a mechanism similar to one proposed for a phonon-driven Floquet matter [32] pump-driven topological phase transition [33]. We demonstrate that, in low dimensional semiconductors, such an exciton-driven Floquet effect can be orders of magnitude larger than the corresponding light-induced Floquet phenomena due to large electron-hole interactions. Within the effective exciton-Floquet Hamiltonian framework, the results of our *ab initio* time-dependent adiabatic GW (TD-aGW) calculations are conceptually and quantitatively explained in terms of the dynamics of the pump-induced exciton polarization which itself plays the role of a driving field.

***Method.*** Our time-resolved, angle-resolved photoemission spectroscopy simulation in a pump-probe setup is shown in Fig. 1 (a). For the computed results presented here for monolayer $MoS_2$, we use a pump light pulse of the form of a truncated half-period sine squared function in time (see supporting information (SI)) with a specific central frequency $\omega_c$ and a duration of 50 fs. Immediately after the pump, we probe the response with a total measure time of 100 fs. In our TD-aGW method [30], we propagate in time the density matrix $\rho(t)$ including electron-hole interactions according to

$$i\frac{d\rho(t)}{dt} = [H(t), \rho(t)], \quad (1)$$

with

$$H(t) = H^{QP} + \delta V^H(t) + \delta \Sigma^{COHSEX}(t) + U^{ext}(t) (2),$$

where $H^{QP}$ is the equilibrium GW quasiparticle Hamiltonian, $\delta V^H$ is the time change in the Hartree energy term from that of equilibrium, $\delta\Sigma^{COHSEX}$ is the time change in self-energy from that of the equilibrium GW self-energy within the static COHSEX approximation which describes the electron-hole interaction in the standard static-screening limit, and $U^{ext}$ is the coupling to the external field (i.e., the pump light). The results for optical response within this formalism in the linear regime of Eq. (1)) is equivalent to the solution of the Bethe-Salpeter equation (BSE) [31,34,35]. The TR-ARPES spectral intensity from a probe pulse measurement with duration of T is computed from our calculation using the lesser Green function $G^<$ [36]

$$I(\omega, T) = -i \, 1/T \int_{T0}^{T0+T} dt_1 \int_{T0}^{T0+T} dt_2 \, G^<(t_1, t_2) e^{i\omega(t_2-t_1)}, (3)$$

where $G^<$ is computed from the density matrix with the $H(t)$ given in Eq. (2) [37] and the lower limit of the integral T0 is the starting time of the probe pulse which is set at the time immediate after the passage of the pump pulse in our simulation. The details of our calculations are given in the SI.

We show in Fig. 1 (b) the time profile of a typical light field with pump frequency $\omega_c$ close to resonance with the energy of an exciton of interest, along with the calculated polarization and the calculated induced self-energy for monolayer MoS$_2$. The time dependence of the polarization P and induced self-energy $\delta\Sigma$ has a direct consequence on the observed ARPES signals: when the pump frequency is close to the excitation energy of an optically bright exciton, the polarization and the corresponding induced self-energy of the system is large and oscillates at the exciton energy. This oscillations in P and $\delta\Sigma$ persist even after the oscillation of the pump pulse decayed, resulting in extended self-energy oscillations that imprint the effects and character of the exciton on the spectral function of the photoexcited hole (i.e., the occupied part of the band structure) and can be observed in ARPES experiments.

**Results.** As a demonstration, we evaluate the TR-ARPES for monolayer MoS$_2$ to take advantage of the large exciton binding energy and optical strength in low dimensional systems, which is expected to yield more distinct exciton features [38-43]. At equilibrium, our GW-BSE calculations [44,45] performed using the BerkeleyGW package [46] give an excitation energy for the lowest optically bright "A" exciton of 1.97 eV and a band gap of 2.48 eV. Fig. 2 (a) shows the computed TR-ARPES with a 1.9 eV pump light and a pulse maximum power intensity of $0.14\times10^{10}$ W/cm$^2$. The observed valence band dispersion agrees well with the equilibrium band structure (that is, with no pump) which is shown by the blue-dashed lines. Besides the valence band signals, the calculations yield additional features, around the K point, located at 0.50 eV below the conduction band minimum (CBM) or at 1.98 eV above the valence band maximum (VBM). Interestingly, instead of a positive band mass associated with the conduction bands, the extra signal at K has a negative effective band mass, which has been suggested to be an ARPES signature of excitons [13-20]. We find that this signal is a satellite feature, replicating the dispersion of the highest valence band around the K point but with a strong **k**-dependent intensity that is largest at the K point, and the energy separation from the VBM (1.98 eV) is virtually identical to the A exciton excitation energy from our direct BSE calculation, indicating that the pump-induced feature is indeed of excitonic origin. To confirm that it is originated from electron-hole interaction, we perform a separate simulation with the same setup but without including any electron-hole interactions. The results are shown in Fig. 2 (b). We clearly see that the extra signal vanishes when the electron-hole interaction is turned off. Since the pump frequency, being 1.9 eV, is off-resonance from the quasiparticle band gap at the K point (which is 2.48 eV), there is negligible conduction band signal. In addition to the ARPES signals from occupied states, we can investigate the unoccupied states (quasi-electron states) by simulating the inverse ARPES. The computed results for inverse ARPES, with and without electron-hole interactions, are shown in Fig. 2(c) and (d), respectively. Similar to the TR-ARPES simulation, we see in the TR-inverse ARPES intensity a satellite of the conduction band bottom separated from the conduction band by the exciton excitation energy in the spectrum that includes electron-hole interactions (Fig. 2(c)), while there are only signals from the quasielectron (i.e., conduction) band states in the inverse ARPES when electron-hole interactions are neglected.

The existence of the valence band satellite feature in the TR-ARPES signals, originating from excitons, and its dispersion can be rationalized in terms of energy conservation: the kinetic energy of the photoemitted electron and the binding energy (measured from the vacuum level) of the *hole* state left behind in the system must add up to either the photon energy or the photon energy *minus* an exciton energy [15]. Hence, in this simple argument, the *dispersion* of the extra hole states near the conduction band is the same as that of *the equilibrium quasiparticle valence band states*. However, this energy argument does not provide any information on the **k**-dependent intensity of the satellite band (which is related to the **k**-space exciton wavefunction) and the argument is only accurate for the dispersion in the weakly pumped regime. Moreover, it cannot predict behaviors in resonantly or moderately to strongly pumped regimes (Fig. 4) that are also readily accessed in contemporary experiments, as we discuss below.

We present here a physical picture for the exciton satellite bands in TR-ARPES in terms of an effective Floquet theory. Given the persistent oscillations of the pump-induced self-energy $\delta\Sigma$ shown in Fig. 1 (b), to illustrate the physics, we construct a two-band framework for the in-gap excitonic features using the highest valence band and the lowest conduction band. For a pump frequency close to the A exciton energy, a well-defined time-dependent polarization induces a time-dependent coupling between these two quasiparticle bands via $\delta\Sigma$, and the oscillation frequency is equal to the A exciton energy. We can write down, after sufficiently time beyond the pump pulse, an effective time-dependent one-particle Hamiltonian,

$$H^{eff}(t) = \sum_k \begin{pmatrix} \epsilon_{vk} + \delta\Sigma_{vvk}(t) & \delta\Sigma_{vck}(t) \\ \delta\Sigma_{cvk}(t) & \epsilon_{ck} + \delta\Sigma_{cck}(t) \end{pmatrix}, (4)$$

where $\varepsilon_{vk}$ and $\varepsilon_{ck}$ are the quasiparticle energy of states in the valence (hole) and conduction (electron) band with momentum **k**, respectively, and $\delta\Sigma(t)$ is the polarization-induced time-dependent self-energy, which includes both the change in the Hartree term and in the static COHSEX self-energy in Eq. (2). (See SI). In particular, the off-diagonal term $\delta\Sigma_{vck}(t)$ is the self-energy oscillation induced by the exciton coherence. For low pump intensity with a frequency close to an exciton resonance that is optically active, we find $\delta\Sigma_{cvk}(t) P_{cvk}(t) e^{-iE^X t}$, where the exciton excitation energy is $E^X$, $P_{cvk}(t)$ is the time-dependent exciton polarization, and both $\delta\Sigma_{vvk}$ and $\delta\Sigma_{cck}$ are negligible at low pump intensity due to their proportionality to excitation densities. The effective Hamiltonian at each wavevector **k** is formally similar to a two-level system coupled by a periodic field, and there is no coupling between **k**-points owing to an assumed spatially uniform pump field which is typical experimentally. We note that the time-dependent self-energy operator $\delta\Sigma(t)$ itself involves a summation over the whole Brillouin zone and its momentum components are related to the envelope function of the exciton (See SI). Solving Eq. (4) within the Floquet formalism, we obtain a discrete series of quasi-energies of exciton-dressed states, which are located at the equilibrium quasiparticle band energy shifted by an integer multiple number of the exciton energy $E^X$. The **k**-resolved quasiparticle spectral function of the lesser Green function for a valence band (in our case the highest valence band) that contributes to the ARPES spectrum, within this picture, can be written as [47-49]

$$I_{vk}(\omega) = \Sigma_m A_k^m \delta(\omega - E_{vk} - mE^X), (5)$$

where $E_{vk}$ (typically close to $\epsilon_{vk}$) is the fundamental quasi-energy of the Floquet Hamiltonian, $m$ is an integer and $A_k^m$ is an amplitude dictated by the **k**-space exciton envelope function. Since the $|m| > 1$ terms are from higher order effects, their spectral intensities are relatively small [47-49]. From Eq. (5), we can immediately see the origin of the satellite bands: they are Floquet satellite bands of the valence electrons due to a time-dependent oscillating self-energy caused by the presence of excitons.

In addition to the prediction of appearance of multiple self-driven exciton-Floquet (satellite) bands and their **k**-dependent spectral intensity, our calculations in the time domain allows us to account for pump-intensity-dependent features in TR-ARPES spectra. As one increases the pump pulse intensity at near resonant frequency, the satellite intensity becomes larger since the system is driven further away from equilibrium, i.e., higher polarization P. This is shown in the comparison between Fig. 3 (a) and (b). In addition to the spectral intensity changes, the energy position of the valence band states is shifted at stronger pump intensity. In the inset of Fig. 3(b), we see that the photoemission peak position of the VBM is shifted down by 30 meV below its equilibrium energy. This is in contrast to the intensity peaks at wavevectors further away from the K point and that of the next valence band below the VBM, which do not show a shift in energy upon increasing pump intensity. This band-selective renormalization of the main quasiparticle peak is due to the fact that the A exciton of monolayer $MoS_2$ is mainly consisted of excitations of free electron-hole configurations from the highest valence band. It is also possible to excite (through selection rules and frequency variations) other exciton states with a pump light of narrow bandwidth which may lead to renormalizations of other bands.

The spectral intensity profile near the energy of the VBM at the K point in Fig. 3(c) summarizes the pump-intensity dependence discussed above. The intensity of the highest valence band signal decreases as we increase the pump intensity, while its peak position shifts to lower energy. At lower intensity, this energy shift in our calculation is linearly proportional to the pump power intensity. On the other hand, the intensity of the satellite signal shown in Fig. 3(d) increases with pump intensity but the position of its peak in energy only shifts slightly relative to the equilibrium VBM. At weaker pump intensity, the satellite peak position (measured from the equilibrium VBM) agrees with the A exciton energy obtained from our equilibrium GW-BSE calculations as indicated by the black-dashed line.

The amount of valence band renormalization (both in spectral weight and peak position) shows a strong **k**-dependence as seen in Fig 3 (b). Such a **k**-dependence can lead to a camel-back feature at higher pump intensity as seen in Fig. 3. In Fig. 4 (a) we show the camel-back dispersion with a pump frequency of 2.0 eV and a pump intensity of $2.25 \times 10^{10}$ W/cm$^2$. Due to the stronger pump intensity and close-to-resonance pump frequency (compared to the results in Fig. 3), the band energy and dispersion as well as spectral weight at the VBM are strongly renormalized while the states at nearby **k** points are less affected. Such a **k**-dependence is caused by the fact that the A exciton has a maximum exciton envelope function amplitude at K [38], and the coupling between interband free electron-hole pairs is strongest for those near this wavevector. At the same time, the dispersion of the satellite band also acquires a camel-back shape, which reflects the corresponding change in the quasiparticle valence band.

The camel-back dispersion can be well reproduced by the effective exciton-Floquet Hamiltonian. In Fig. 4 (b), we plot the quasi-energy obtained from the Floquet calculations at selected **k** points. We emphasize here that the green dots in Fig. 4(b) indicate the value of the Floquet quasi-energies *only* regardless of its spectral weight, whereas the TR-ARPES results from our TD-aGW calculations show both the energy and weight of the spectral function at a given **k** through the color intensity scale. We clearly see that the quasi-energy dispersion agrees well with the TD-aGW spectral intensity peak dispersion of both the highest valence band and its satellite, which validates the physical picture presented above. (The band of green dots near 0.5 eV in Fig. 4(b) is a Floquet m=-1 satellite band associated with the conduction band which can only be seen in an inverse TR-ARPES measurement.)

Such camel-back feature has been seen in earlier work [26] and was interpreted as signature of an out-of-equilibrium exciton insulator state created by the optical pump. This out-of-equilibrium exciton insulator proposal originates from an expected camel-back-like renormalization of the VBM and CBM in traditional excitonic insulators, in which the exciton binding energy is comparable to or larger than the quasiparticle bandgap of the normal band insulator phase [7]. (This is not true for the present case of a monolayer $MoS_2$. [38]) Here we show that camel-back band renormalization in fact arises from a self-driven exciton-Floquet physics, and that this effect can be driven from arbitrary bands and with excitons with either zero or finite center-of-mass (COM) momentum. More generally, excitons generated indirectly through some relaxation process or by means other than light may also lead to such renormalizations, as is the case of spin-dark or finite COM-momentum excitons. This is a main difference in the self-driven exciton-Floquet picture with respect to the traditional external light-driven Bloch-Floquet states: in the former, the renormalization only occurs for specific states that couple to the oscillating exciton field; in the latter, replica states emerge throughout the BZ, as long as there are sizeable dipole matrix elements coupling the states to nearby bands.

One can show that in general the band renormalization $\Delta$ obtained from a Floquet mechanism is proportional to the coupling strength squared and inversely proportional to the detuning $\delta$. For the light-driven Bloch-Floquet state, we obtain $\Delta_{\boldsymbol{k}}^{\text{light}} \sim \frac{|d_{cv\boldsymbol{k}}|^2}{\delta}$, where $d_{cv\boldsymbol{k}}$ is the optical matrix elements while for exciton-Floquet we have $\Delta_{\boldsymbol{k}}^{\text{xct}} \sim \frac{|\delta\Sigma_{cv\boldsymbol{k}}|^2}{E_b} \sim E_b A_{cv\boldsymbol{k}}^2 n_{\text{xct}}$, where $E_b$ is the exciton binding energy, $A_{cv\boldsymbol{k}}$ is the exciton envelop function and $n_{\text{xct}}$ is the exciton density. We find that, with a photon energy of 2.2 eV and an external field amplitude of $2 \times 10^{-4}$ a.u, the band renormalization from a light-driven Bloch-Floquet is about 65 times smaller than that from an exciton-Floquet in MoS$_2$, which makes low-dimensional semiconductors an excellent platform to study Floquet phenomena.

By analyzing the time snapshot of the Hamiltonian from our TD-aGW results, we find that both the diagonal ($\delta\Sigma_{vvk}$, $\delta\Sigma_{cck}$) and the off-diagonal ($\delta\Sigma_{vck}$ self-energy matrix elements in Eq. (4) are of the same order of magnitude under the stronger pump intensity corresponding to Fig. 4(b). However, they modify the band gap in opposite directions. The diagonal term shrinks the band gap, while the off-diagonal term leads to a band repulsion. We find that the resulting quasiparticle band gap is increased at both 1.9 eV and 2.0 eV pump frequency considered. However, a dramatic change in band renormalization occurs when the pump frequency is close to the equilibrium interband free electron-hole transition energy, which is at 2.48 eV. At both 2.4 eV and 2.6 eV pump photon energy, the calculated spectral peak position corresponding to the valence band states instead shifts upwards in energy (i.e., lowering the energy for the excited holes). This happens because, with the pump frequency being close to the electron-hole continuum, the off-diagonal terms of $\delta\Sigma_{cv}$ at different $\mathbf{k}$ no longer oscillate at a single frequency (see SI), so the overall band repulsion effects are washed out due to the destructive interference of different frequencies of the polarization oscillations.

In conclusion, we have shown from first-principles time-dependent calculations [30] that remarkable excitonic features, containing quantitative information on the exciton energy and wavefunction, can be observed in TR-ARPES experiments with a pump photon energy close to exciton excitation energies under achievable experimental conditions. We demonstrate that such effects can be significantly stronger than light-induced Floquet on monolayer MoS$_2$, and, with increasing pump intensity, we observe larger $\mathbf{k}$-selective band renormalization (both in energy and dispersion). Our first-principles calculations show that exciton-driven Floquet effects, first theorized in model Hamiltonian calculations [26,33], are greatly enhanced in low-dimensional materials resulting in a concomitant enhancement of the many-electron interactions. Our results conclusively demonstrate the signatures of excitons and the quantitative information that may be extracted from them in TR-ARPES experiments and open a new avenue for the accurate study of the properties of excitons, in particular their $\mathbf{k}$-space wavefunctions, through careful theoretical and experimental analyses of TR-ARPES.

This work was supported by the Center for Computational Study of Excited State Phenomena in Energy Materials (C2SEPEM), which is funded by the U.S. Department of Energy, Office of Science, Basic Energy Sciences, Materials Sciences and Engineering Division under Contract No. DE-AC02-05CH11231, as part of the Computational Materials Sciences Program. YHC, DYQ and FHJ thank Keshav Dani for helpful discussion. YHC thanks Zhenglu Li and Yi Lin for helpful discussion. We acknowledge the use of computational resources at the National Energy Research Scientific Computing

Center (NERSC), a DOE Office of Science User Facility supported by the Office of Science of the U.S. Department of Energy under Contract No.DE-AC02-05CH11231.

## Figures

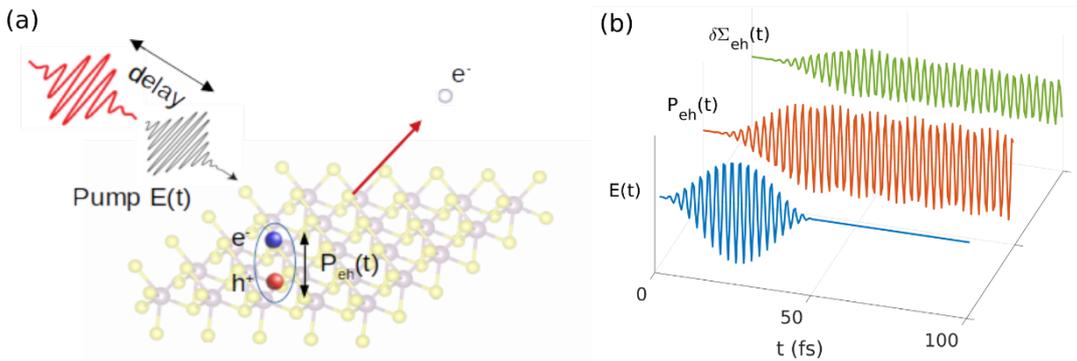

FIG. 1. (a) Sketch of the pump-probe setup and the physical picture of exciton polarization induced TR-ARPES signals. (b) Time dependence of the external pump field E(t), calculated exciton polarization $P_{eh}(t)$, and calculated induced self-energy $\delta\Sigma(t)$, for a monolayer $MoS_2$.

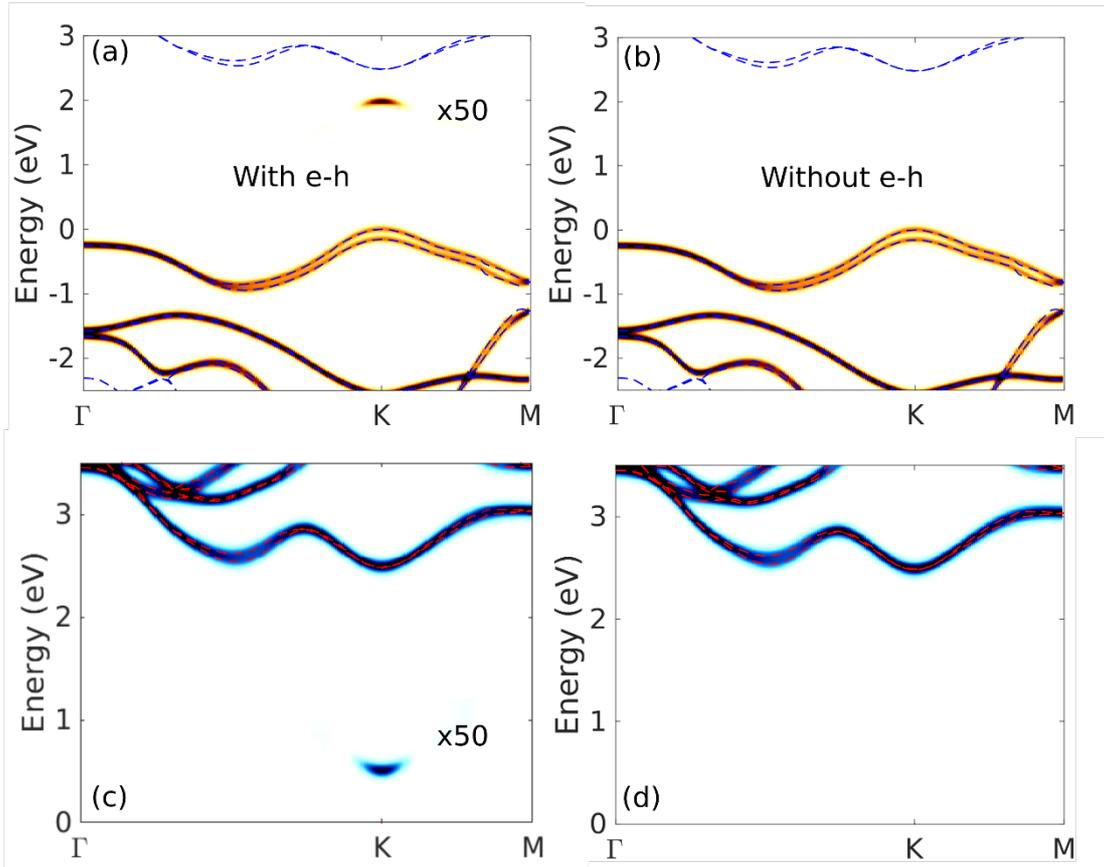

FIG. 2. Computed TR-ARPES intensity of monolayer $MoS_2$: (a) with electron-hole interaction and (b) without electron-hole interaction with a pump frequency of 1.9 eV of 50 fs duration and a maximum pump power intensity of $0.14 \times 10^{10}$ W/cm². The spectral intensity near the conduction band is enhanced by 50 times for better visibility. Computed inverse ARPES intensity in a similar pump setup: (c) with e-h interaction and (d) without e-h interaction. The zero of the energy is set at equilibrium valence band maximum.

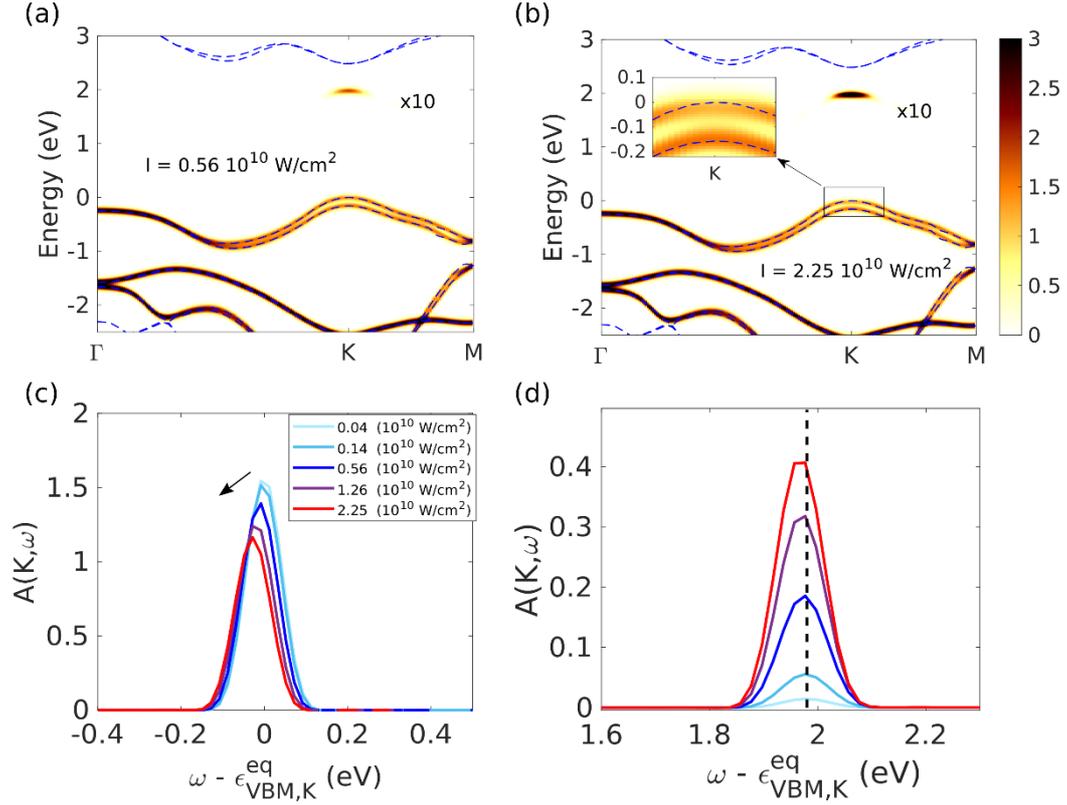

FIG. 3. Simulated TR-ARPES with 1.9 eV pump frequency and (a) maximum intensity of $0.56\times10^{10}$ W/cm² and (b) of $2.25\times10^{10}$ W/cm². Inset in (b) shows the quasiparticle band renormalization around VBM. The spectral intensity above 1 eV is increased by 10 times for visibility. Intensity dependence of spectral function at K for (c) VBM (near 0 eV) and (d) satellite feature (near 2 eV). Black dashed line in (d) indicates the equilibrium A exciton excitation energy from GW-BSE calculation.

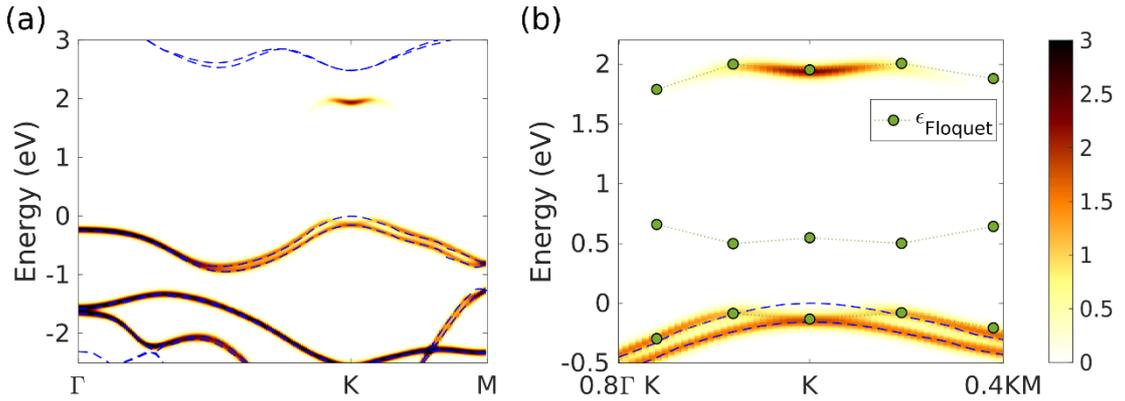

FIG. 4. (a) Camel-back features in the simulated TR-ARPES with pump frequency of 2.0 eV and intensity of $2.25\times10^{10}$ W/cm².The spectral intensity above 1 eV is increased by a factor of 2 for visibility. (b) Comparison between the spectral intensity in the region around the VBM from (a) and the quasi-energy bands (green dots) solved from the effective exciton-Floquet Hamiltonian. Blue-dashed lines are the equilibrium bands. The green dots indicate the value of the Floquet quasi-energies *only* (regardless of their spectral weight), whereas the TR-ARPES results from TD-aGW show both the energy and weight of the spectral function at a given **k** through the color intensity scale. The band of green dots near 0.5 eV is a Floquet m=-1 satellite band associated with the conduction band which should not be visible in TR-ARPES measurements.